\documentclass[journal=jacsat,manuscript=article]{achemso}
\usepackage{dcolumn}
\usepackage{bm}
\usepackage[mathlines]{lineno}

\usepackage[hidelinks,colorlinks=true,citecolor=blue,linkcolor=red]{hyperref}

\usepackage{chemformula} 
\usepackage[T1]{fontenc}
\usepackage{textcomp}
\usepackage{units}
\usepackage{hyperref}
\usepackage{amsmath}
\usepackage{multirow}
\usepackage{rotating}
\usepackage{graphicx}
\usepackage{amsmath}
\usepackage{wasysym}
\usepackage{float}
\usepackage{color}
\usepackage{miller} 
\usepackage[english]{babel}
\usepackage{upgreek}
\usepackage[all]{hypcap}
\usepackage{xcolor}
\usepackage{comment}
\usepackage{units}
\usepackage{textcomp}
\usepackage{natbib}

\author{Chitran Ghosal}
\affiliation{Solid Surface Analysis, Institute of  Physics, Chemnitz University of Technology, 09126 Chemnitz, Germany}

\author{Siheon Ryee}
\email{sryee@physnet.uni-hamburg.de} 
\affiliation{I. Institute of Theoretical Physics, University of Hamburg, Notkestrasse 9, 22607 Hamburg, Germany}

\author{Zamin Mamiyev}
\affiliation{Solid Surface Analysis, Institute of  Physics, Chemnitz University of Technology, 09126 Chemnitz, Germany}

\author{Niklas Witt}
\affiliation{Institute for Theoretical Physics and Astrophysics, University of Würzburg, Am Hubland, 97074 W\"urzburg, Germany}
\alsoaffiliation{I. Institute of Theoretical Physics, University of Hamburg, Notkestrasse 9, 22607 Hamburg, Germany}
\alsoaffiliation{The Hamburg Centre for Ultrafast Imaging, Luruper Chaussee 149, 22761 Hamburg, Germany}

\author{Tim O. Wehling}
\affiliation{I. Institute of Theoretical Physics, University of Hamburg, Notkestrasse 9, 22607 Hamburg, Germany}
\alsoaffiliation{The Hamburg Centre for Ultrafast Imaging, Luruper Chaussee 149, 22761 Hamburg, Germany}

\author{Christoph Tegenkamp}
\email{christoph.tegenkamp@physik.tu-chemnitz.de}
\affiliation{Solid Surface Analysis, Institute of  Physics, Chemnitz University of Technology, 09126 Chemnitz, Germany}

\title{Electronic correlations in epitaxial graphene: Mott states proximitized to a relativistic electron gas}


\begin{document}
\begin{abstract}
Graphene, renowned for its exceptional electronic and optical properties as a robust 2D material, traditionally lacks electronic correlation effects. Proximity coupling offers a promising method to endow quantum materials with novel properties. In this study, we achieve such a proximity coupling by intercalating Sn between the buffer layer of graphene on SiC(0001), allowing us to explore the coupling between a correlated 2D electron gas and a Dirac metal. This results in the stabilization of Sn-$\sqrt{3}$ superlattice structures at the interface, which reveal Mott-Hubbard bands, in excellent agreement with both experimental observations and theoretical predictions. Additionally, we found signatures of quasiparticle peaks close to the Fermi energy, in detail depending on the hybridization strength and doping level. 

\end{abstract}

\section{Introduction}

Strong electronic correlation effects are prevalent in low-dimensional systems due to reduced Coulomb screening and the low kinetic energy of charge carriers. Thereby, two-dimensional (2D) materials play a special role in this context, as different stacking sequences may come along with new quantum states \cite{Guo2021, Cao2018, Cao_2018_2, Tang2020, Tang2023, Nuckolls2024, Bernevig2024}.

A central concept in stacked systems are moiré structures, which have demonstrated superconducting, correlated insulating, and topological behavior, such as in twisted bilayer graphene \cite{Cao2018,Cao_2018_2}. This controlled symmetry breaking and formation of flatbands can be extended to multilayer systems, including rotated graphene bilayer systems \cite{He2021}.
Transition metal dichalcogenides (TMDs), inherently 3-layer systems, also fit into this category. For example, the preserved inversion symmetry in 1T-TaS$_2$  results in a insulating charge density wave system, while the inversion-broken 2H-TaS$_2$ 2D layer exhibits Ising superconductivity \cite{Lu2015,Law2017}. The Fermi-surface nesting in these systems is highly sensitive to doping. Instead of an antiferromagnetic ground state, quantum fluctuations prevent magnetic order, leading to spin-liquid behavior \cite{Law2017}. Additionally, stacks of TMDs, such as 4Hb-TaS$_2$ and heterogeneous Ta-dichalcogenide bilayers, display a delicate interplay of correlated insulator, charge transfer and heavy fermion physics~\cite{ruan_evidence_2021,vano_artificial_2021,crippa2024} leading to signatures of exotic quantum states like chiral superconductivity \cite{Silber2024}.
Furthermore, interlayer correlation effects and charge transfer are critical for the superconductivity observed in bilayer nickelates \cite{Sun2023,Ryee2024}. 

The overarching theme of these correlated states is driven by Mott physics. The physics of proximity coupling in these systems can be successfully explained using extended Hubbard and heavy fermion models, which describe insulating Mott- and Wigner-Mott systems and consider the roles of spin and charge order as well as band filling factors \cite{Regan2020,Shimazaki2020,Li2021,Ghiotto2021,Tang2023}.

The interfaces of the aforementioned systems play a crucial role in realizing modulated periodic potentials. Epitaxial growth is a well-established method for achieving extremely clean and atomically well-defined interfaces. For example, epitaxial graphene can be grown on large scales and is suitable for quantum metrology \cite{Emtsev2009, Kruskopf2016, Yin2024}.
Moreover, the sp$^2$-hybridized sheet of C-atoms on SiC(0001) is ideal for subsequent intercalation. By using various elements for intercalation between the buffer layer (BL) and the SiC surface, the 2D carbon layer is lifted, revealing the honeycomb structure and the Dirac cone. Intercalation has been used to control interface charges, enabling charge-neutral, p-type, or n-type graphene (e.g., H \cite{Riedl2009}, Ge \cite{Emstev2011}, Au \cite{Gierz2010}, S \cite{Wolff2024}, Pb \cite{Schaedlich2023}).
Recently, extreme doping schemes with lanthanides, such as Gd, Yb, and Tb, have garnered research interest due to their ability to enforce correlation effects in monolayer graphene. These effects include band renormalization \cite{Link2019}, Lifshitz transitions \cite{Rosenzweig2019}, and Kohn-Luttinger superconductivity \cite{Herrera2024}.

Among these, group-IV elements adsorbed on semiconducting surfaces have attracted attention for their strong spin–orbit coupling and correlated electronic behavior. Phenomena such as the Mott insulating state, chiral superconductivity, and topological edge states have been observed on these surfaces \cite{Hansmann2013, Jaeger2018, Nakamura2018, Cao2018a, Wu2020, Ming2023}. Submonolayer coverages of Sn on SiC(0001) have been shown to host a robust Mott state \cite{Glass2015}.
Thus, the intercalation of Sn into BL/SiC structures is a promising prototype system for bringing a correlated state into proximity with a highly mobile electron gas. Recent studies using plasmon spectroscopy have examined Sn intercalation in greater detail. A fully saturated interface layer results in quasi-free and charge-neutral graphene, while Sn submonolayer structures produce n-type doped graphene \cite{Mamiyev2022, Mamiyev2024}.

In this paper, we demonstrate that the proximity coupling of an electronically correlated 2D electron gas to Dirac electrons in graphene reveals signatures of both a Dirac metal and strongly correlated Hubbard bands. Using scanning tunneling microscopy and spectroscopy, supplemented by electron diffraction and electron energy loss spectroscopy, we identify the Mott-Hubbard bands from the interface layer and observe renormalization effects on the Dirac states. Through dynamical mean-field theory (DMFT) calculations, we
reveal that besides the hybridization and Coulomb interaction also the charge transfer plays an important role for electronic states that emerge in such correlated systems.

\section{Materials and Methods}
\textit{Phase preparation and analysis: }
The experiments were performed  in various ultra-high vacuum (UHV) systems, all operating at base pressures of 5 $\times$ 10$^{-9}$~Pa. Samples were transferred between systems without breaking the vacuum using a UHV suitcase. The Sn intercalation process and interface structures were controlled and studied with a SPA-LEED (spot profile analysis low energy electron diffraction) system. Plasmonic and interband excitations were measured using high-resolution electron energy loss spectroscopy (EELS-LEED), providing high energy and momentum resolution, typically around 25 meV and 0.001~\AA$^{-1}$ \cite{Scheithauer1986,Mamiyev2021a}. Scanning tunneling microscopy (STM) experiments were performed at 4 K using W-tips, and scanning tunneling spectroscopy (STS) was conducted with a lock-in technique (4 mV, 800 Hz).

Buffer layer (BL) graphene samples were grown epitaxially by Si sublimation on the Si face of 6H-SiC(0001) substrates as described in detail in Ref.~\cite{Emtsev2009}. For the intercalation experiments, Sn was evaporated at room temperature from a Mo-crucible at a deposition rate of 0.15 monolayer/min, followed by several annealing steps until LEED reveals $\sqrt{3} \times \sqrt{3}$ spots as a hallmark for a Mott phase on SiC(0001)\cite{Glass2015}. The Sn coverage was checked using a quartz microbalance and calibrated via the $\alpha$-Sn reconstruction on Si(111) as outlined in Ref.~\cite{Mamiyev2022}. Temperatures were measured using an infrared pyrometer (Impac IGA).

\begin{figure}[t]
\begin{center}
	\includegraphics[width=.6\columnwidth]{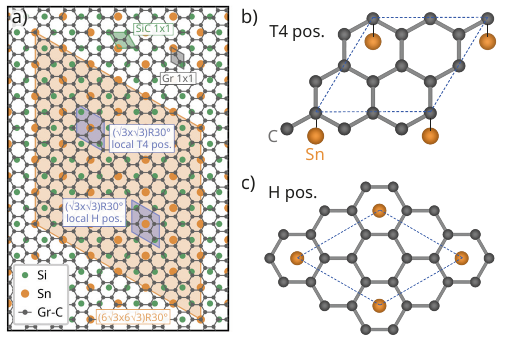}
	\caption{(a): Illustration of the full $6\sqrt{3}\times6\sqrt{3}$ supercell reconstruction of graphene on SiC(0001) and $\sqrt{3}\times\sqrt{3}$ supercells capturing the Sn superstructure with different local alignments of approximate $2\times2$ graphene sheets with respect to SiC. The unit cells of graphene and SiC are also shown. For the SiC substrate, only the topmost Si atom is depicted. (b,c): Crystal structure of the approximate $\sqrt{3}\times\sqrt{3}$ cell model for EG/Sn in the stacking configuration of Sn being in the T4 position or in the center of the hexagon (H). The unit cell is highlighted with blue dashed lines and the SiC substrate is not shown.   \label{FIG0}}
\end{center}
\end{figure}

\textit{Electronic structure calculations}:
To describe the electronic structure of graphene in proximity to a Sn layer with ($\sqrt{3} \times \sqrt{3}$) symmetry on the SiC substrate, we perform density functional (DFT) calculations of the layered heterostructure. We employ the Vienna ab initio software package (VASP)~\cite{Kresse1993,Kresse1996,Kresse1996a} within the generalized gradient approximation of Perdew, Burke, and Ernzerhof (GGA-PBE)~\cite{Perdew1996} and the projector augmented wave (PAW)~\cite{Bloechl1994,Kresse1999} basis sets. We use a $12\times12\times1$ $k$-mesh, a plane wave cutoff of $400$~eV, and DFT-D3 van der Waals corrections~\cite{Grimme2010} in our calculations. We relaxed the distance between the Sn layer and graphene using conjugate
gradient algorithm for ionic relaxation until all force components are smaller than
$0.005$~eV\AA$^{-1}$. The data of the DFT calculations is made available at~\cite{DFTNOMAD}.

In our calculation we address a simplified structure to account for the main experimental findings. Instead of the rather large supercell associated with $6\sqrt{3}\times6\sqrt{3}$ reconstruction of graphene on the SiC(0001) substrate~\cite{Riedl2010}, we use a smaller $\sqrt{3}\times\sqrt{3}$ cell in our DFT calculations~\cite{Mattausch2007} which fully contains the Sn-($\sqrt{3} \times \sqrt{3}$) triangular lattice reconstruction \cite{Glass2015}. As is seen in Fig.~\ref{FIG0}a), this ($\sqrt{3} \times \sqrt{3}$) supercell approximates local registries of the full $6\sqrt{3}\times6\sqrt{3}$ supercell with varying stacking configuration of the graphene sheet relative to the SiC surface. In our approximate supercell, graphene is stretched to fit in a $2\times2$ arrangement on SiC. We consider two limiting cases of the stacking configuration: the Sn atoms are located either right below a C atom [Fig.~\ref{FIG0}b), ``T4'' position] or below the center of the hexagon [Fig.~\ref{FIG0}c), ``H'' position]. They realize the maximal and the minimal hybridization between Sn and C atoms, respectively.

\textit{Model Hamiltonian}:
For studying the strongly correlated nature of $p_z$ electrons stemming from Sn-($\sqrt{3} \times \sqrt{3}$) on SiC(0001) \cite{Glass2015}, we consider an effective low-energy model with Hamiltonian $\mathcal{H} = H_0 + H_\mathrm{int}$ capturing the DFT band structure (c.f.~Figure~\ref{fig_t1} below). In the kinetic part $H_0$ of our model, we only consider contributions by Sn and graphene $p_z$ orbitals as:
\begin{align} \label{eq:model}
H_0 = \sum_{i\neq j,\sigma}(t^\mathrm{C}_{ij} c^\dagger_{i\sigma} c_{j\sigma} + t^\mathrm{Sn}_{ij} d^\dagger_{i\sigma} d_{j\sigma}) + \sum_{i} ( \epsilon_\mathrm{C} n^c_{i} + \epsilon_\mathrm{Sn} n^d_{i} ) + \sum_{i\sigma}V_{ij}(c^\dagger_{i\sigma} d_{j\sigma} + \mathrm{h.c.})\,.
\end{align}
The indices $i$ and $j$ refer to the in-plane lattice sites and $\sigma$ ($\sigma \in \{ \uparrow, \downarrow \}$) denotes the electron spin. $c^\dagger_{i\sigma}$ ($c_{i\sigma}$) is the electron creation (annihilation) operator for graphene and $d^\dagger_{i\sigma}$ ($d_{i\sigma}$) for Sn. The corresponding number operators are $n^c_i = \sum_{\sigma} n^c_{i\sigma} = \sum_{\sigma} c^\dagger_{i\sigma}c_{i\sigma}$ and $n^d_i = \sum_{\sigma} n^d_{i\sigma} = \sum_{\sigma} d^\dagger_{i\sigma}d_{i\sigma}$, respectively. 
$t^\mathrm{C/Sn}_{ij}$ is the hopping amplitude between $i$ and $j$ sites. We hereafter simply denote the $n$-th neighbor hopping by $t^\mathrm{C/Sn}_{n}$. For C sites (graphene), we use only the nearest-neighbor hopping $t^\mathrm{C}_1 = -2.8$~eV which is the typical value for the pristine graphene \cite{wehling2011}. For Sn sites, we set $t^\mathrm{Sn}_1 = 27.3$~meV used in Ref.~\cite{Glass2015} and $t^\mathrm{Sn}_2 = -0.3881\, t^\mathrm{Sn}_1 $ and $t^\mathrm{Sn}_3 = 0.1444\, t^\mathrm{Sn}_1 $ for longer-range hopping by referring to the {\it ab initio} data for Sn-($\sqrt{3} \times \sqrt{3}$) on Si(111) \cite{Li2013,Jaeger2018}. 
$\epsilon_\mathrm{C/Sn}$ is the on-site energy level with respect to the chemical potential. The last term in Eq.~(\ref{eq:model}) couples graphene and Sn-($\sqrt{3} \times \sqrt{3}$) via $V_{ij}$, which accounts for the hybridization between Sn and C atoms.  To model $V_{ij}$, we adopt an approximation used for parametrizing interlayer couplings in the twisted bilayer graphene \cite{trambly_2010,moon2013}, which reads
\begin{align}
	V_{ij}=V_0 \mathrm{exp}\Bigg( - \frac{\sqrt{l_{ij}^2 + l_\perp^2} - l_\perp}{\delta_0} \Bigg) \frac{l_\perp^2}{(l_{ij}^2 + l_\perp^2)} ,
\end{align}
where $V_0$ is the hybridization strength between vertically located Sn and C atoms, $l_{ij}$ the in-plane distance between atoms at sites $i$ and $j$, $l_\perp=3.45$~\AA~the DFT-optimized vertical spacing between the Sn layer and graphene, and $\delta_0=0.184\,a$ the decay length ($a$ is the distance between the neighboring C atoms in graphene). Thus we have only a single free parameter $V_0$.  For simplicity, we truncate $V_{ij}$ for $l_{ij} > a$ since $V_{ij}$ decays exponentially with $l_{ij}$. We use $V_0$ and the crystal field $\Delta \equiv \epsilon_\mathrm{Sn}- \epsilon_\mathrm{C}$ as our control variables. 

To account for correlation effects, we use the Hubbard interaction for $H_\mathrm{int}$ which reads
\begin{align}
H_\mathrm{int} = U\sum_{i} \big( n^d_{i\uparrow} - \langle n^d_{i\uparrow} \rangle_0 \big) 
\big( n^d_{i\downarrow} - \langle n^d_{i\downarrow} \rangle_0 \big),
\end{align}
Here, $U$ denotes the local Coulomb (Hubbard) interaction for electrons sitting on the same Sn site, and $\langle n^d_{i\sigma} \rangle_0$ is the electron occupation in the noninteracting case ($U=0$). The above expression for $H_\mathrm{int}$ implies that our $H_0$ already includes the static mean-field self-energy \`a la the DFT, and thus $H_\mathrm{int}$ accounts for only dynamical correlations. 

\textit{DMFT calculations:}
The Hamiltonian $\mathcal{H}$ is solved using the DMFT to address nonperturbatively local quantum fluctuations~\cite{DMFT}. 
A numerically exact hybridization-expansion continuous-time quantum Monte Carlo algorithm is used as an impurity solver~\cite{CTQMC,comctqmc}. We restrict ourselves to paramagnetic phases without any spatial symmetry-breaking, and we set the temperature to $T = 0.005~\mathrm{eV} \simeq 58~\mathrm{K}$. We use $U=1.2$~eV as a representative value considering the previous studies on Sn surface systems \cite{Glass2015}. To calculate the momentum-dependent spectral functions $A(\mathbf{k},\omega)$ ($\mathbf{k}$: crystal momentum, $\omega$: real frequency) and the momentum-integrated spectral function $A(\omega)$, we analytically continue the DMFT local self-energy $\Sigma(i\omega_n)$ ($\omega_n$: fermionic Matsubara frequency) defined on the Matsubara frequency axis to the real frequency axis by employing the maximum entropy method \cite{Jarrel,Bergeron}. The chemical potential is adjusted during the DMFT self-consistency to ensure the average electron occupation $\langle n \rangle$ of $\langle n \rangle = 9$ per unit cell, with 8 electrons from the 8 carbon and one electron from the Sn atom per unit cell. 

\section{Results and Discussion}
\subsection{Structure of EG/Sn interface: SPA-LEED and STM}
\begin{figure}[t]
\begin{center}
	\includegraphics[width=1\columnwidth]{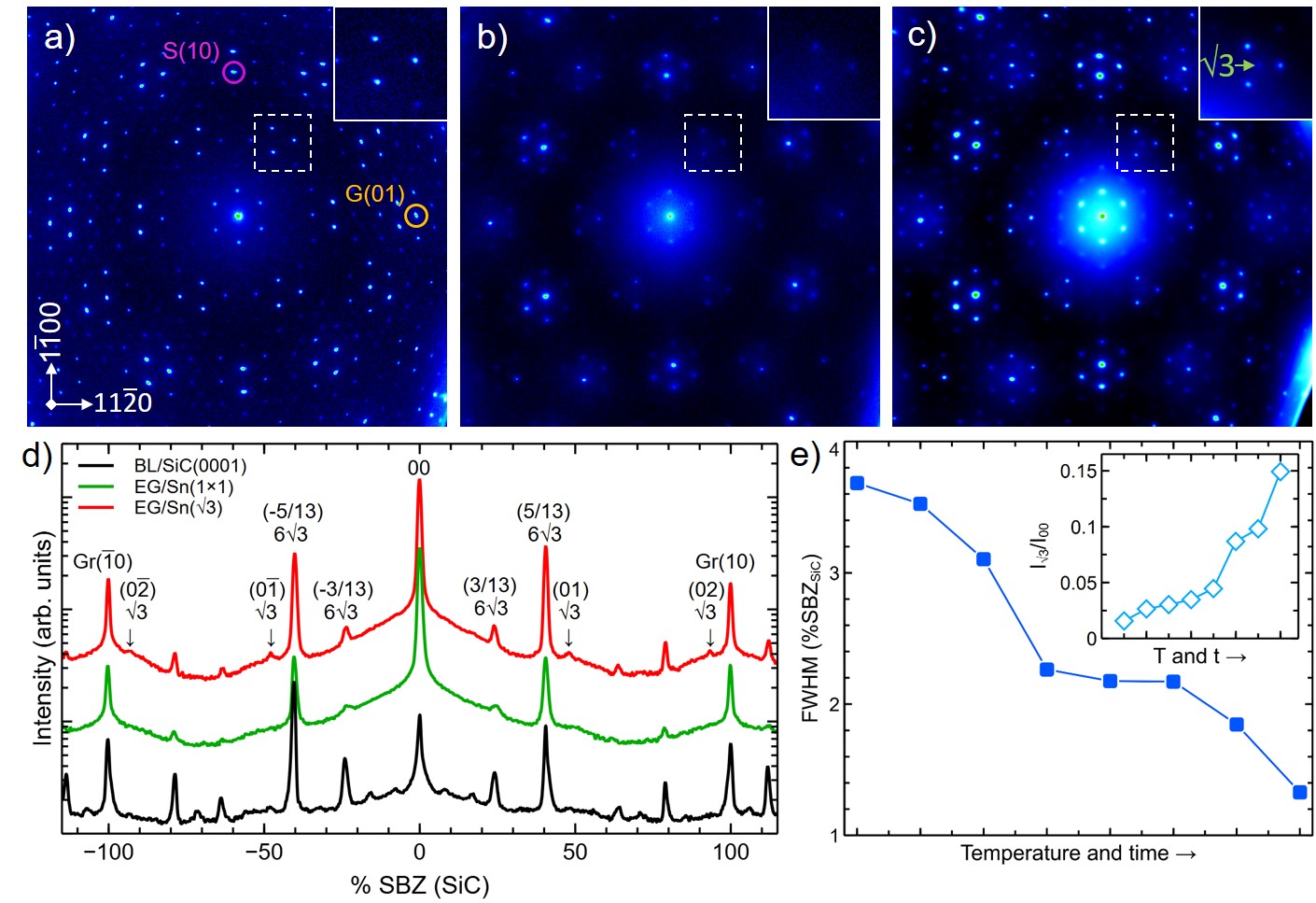}
	\caption{(a) SPA-LEED image of the buffer layer (BL) before intercalation of Sn. (b) SPA-LEED image after intercalation of a full monolayer of Sn. (c) LEED image after de-intercalation of Sn revealing ($\sqrt{3} \times \sqrt{3}$) reflexes. All LEED images were taken at $E=150$~eV and 300~K.  (d) Line scans of all three phases taken along the $[11\overline{2}0]$-direction. (e) FWHM and normalized  intensity (inset) of a ($\sqrt{3} \times \sqrt{3}$)-spot after different cycles of annealing. \label{FIG1}}
\end{center}
\end{figure}

The intercalation process of Sn and formation of the $\sqrt{3} \times \sqrt{3}$-phase was monitored by SPA-LEED and STM. In Fig.~\ref{FIG1} we show a sequence of diffraction images before and after intercalation as well partial de-intercalation of Sn. Starting with the BL/SiC(0001), the LEED pattern reveals all characteristic spots of the $6\sqrt{3} \times 6\sqrt{3}$ reconstruction of the carbon honeycomb structure with the SiC(0001) surface \cite{seyller2008epitaxial,Riedl2010}.   
By intercalation of Sn the BL is transformed into a quasi free monolayer graphene. We adsorbed 5 ML of Sn on BL/SiC(0001) at room temperature and annealed the film subsequently at 1070~K for 15 minutes \cite{Mamiyev2022}. The LEED pattern in panel b) revealed the same symmetry, but shows in addition the so-called bell shape background at all graphene integer spots, which is a hallmark for the formation of free-standing graphene \cite{Petrovic2021}. In agreement with previous measurements and also the STM data, this phase refers to an intercalated  Sn-(1$\times$1) phase. Although crystalline Sn-phases are formed only locally, the dangling bonds of the SiC substrate are on average saturated coming along with charge neutral graphene, e.g., seen in ARPES measurements \cite{Federl2024}.
Since the formation of a Sn-($\sqrt{3} \times \sqrt{3}$) phase by direct intercalation of 1/3 of a monolayer of Sn did not succeed so far, we started out with the fully saturated interface phase. Annealing the surface to 1320~K for 5 minutes came along with the emergence  of ($\sqrt{3} \times \sqrt{3}$)-reflexes in the SPA-LEED pattern, highlighted in the inset of Fig.~\ref{FIG1}c). Since this phase is formed by de-intercalation, the domains are limited in size but can be enlarged by subsequent annealing cycles, i.e., the full width at half maximum (FWHM) of the $\sqrt{3}$-reflexes decreases while  normalized integral intensity increases as shown in Fig.~\ref{FIG1}e). From the analysis of the FWHM the  domain sizes vary between 20~nm and 80~nm.

\begin{figure}[t]
	\begin{center}
	\includegraphics[width= \linewidth]{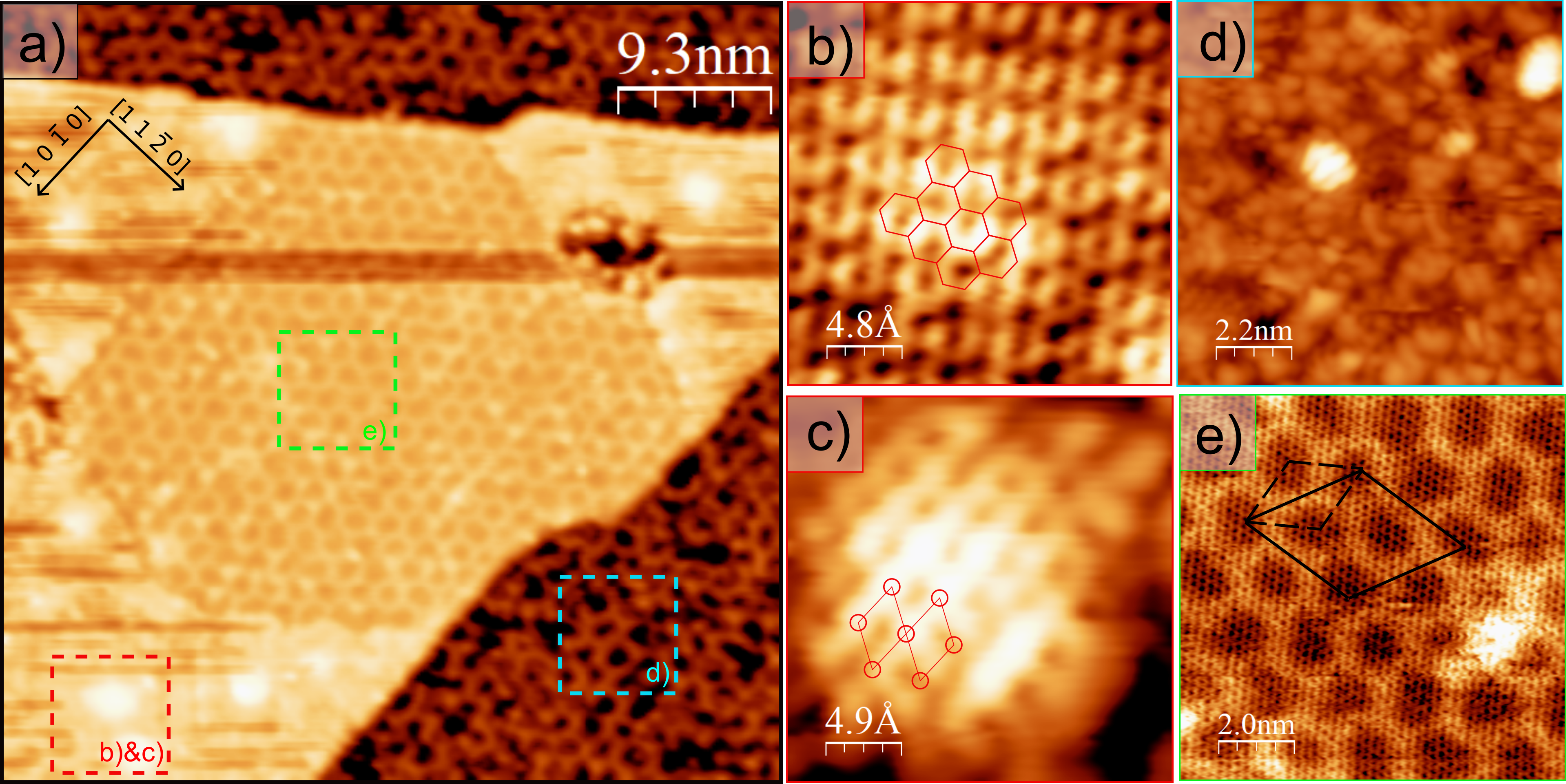}
	\caption{(a) Large scale STM image (+2.0~V, 0.5~nA) showing  three different phases, i.e., Sn-(1$\times$1) (red square,(zoom in shown in (b) and (c))), disordered BL (blue square,(d)) and the Sn-$\sqrt{3}$  (green square,(e)). (b,c) Atomically resolved STM of the Sn-(1$\times$1) phase at 100~mV, 25~pA (b) and 1~V, 350~pA (c) showing the graphene lattice and interface structure, respectively. d) Disordered phase measured at +1.5~V, 500~pA. (e) Sn-$\sqrt{3}$ phase (V=+0.5~V, 500~pA) showing both the graphene lattice and a (6$\times$6) periodicity (dashed rhombus). The solid rhombus marks the ($6\sqrt{3} \times 6\sqrt{3}$) unit cell.}\label{FIG2}
	\end{center}
\end{figure}

In Fig.~\ref{FIG2}a)  we show a typical large scale STM image after the Sn monolayer intercalation and partial de-intercalation. As obvious, the surface is inhomogeneous  and mainly three different phases can be identified. 
Depending on the bias voltage, in panel b) and c) and either the graphene lattice and atomic structure of the interface are shown. This phase refers to the fully intercalated interface structure and shows partially a Sn-(1$\times$1) reconstruction. The proximity of these crystalline Sn islands to the graphene layer induces the intervalley scattering, thus the graphene shows a ($\sqrt{3} \times \sqrt{3}$)-reconstruction \cite{Mamiyev2024b}. 
In panel d) we show an apparently more or less fully de-intercalated area. Irrespective of the STM bias voltages, we were not able to resolve any periodic lattice structures, thus this phase is reminiscent of a disordered BL.  The most important phase for the further discussion is shown in panel e). Besides   the graphene lattice,  the (6$\times$6) symmetry of the former BL is seen. We will show in the following that this phase contains locally Sn-$\sqrt{3}$ phases, which  are responsible for the additional LEED reflexes seen in Fig.~\ref{FIG1}c).  In contrast to the phase shown in panel b), the graphene itself does not reveal any reconstruction. For this phase the interface layer is obviously not enabling intervalley scattering within the graphene layer. 
  
\begin{figure}[t]
		\centering
		\includegraphics[width= .4\linewidth]{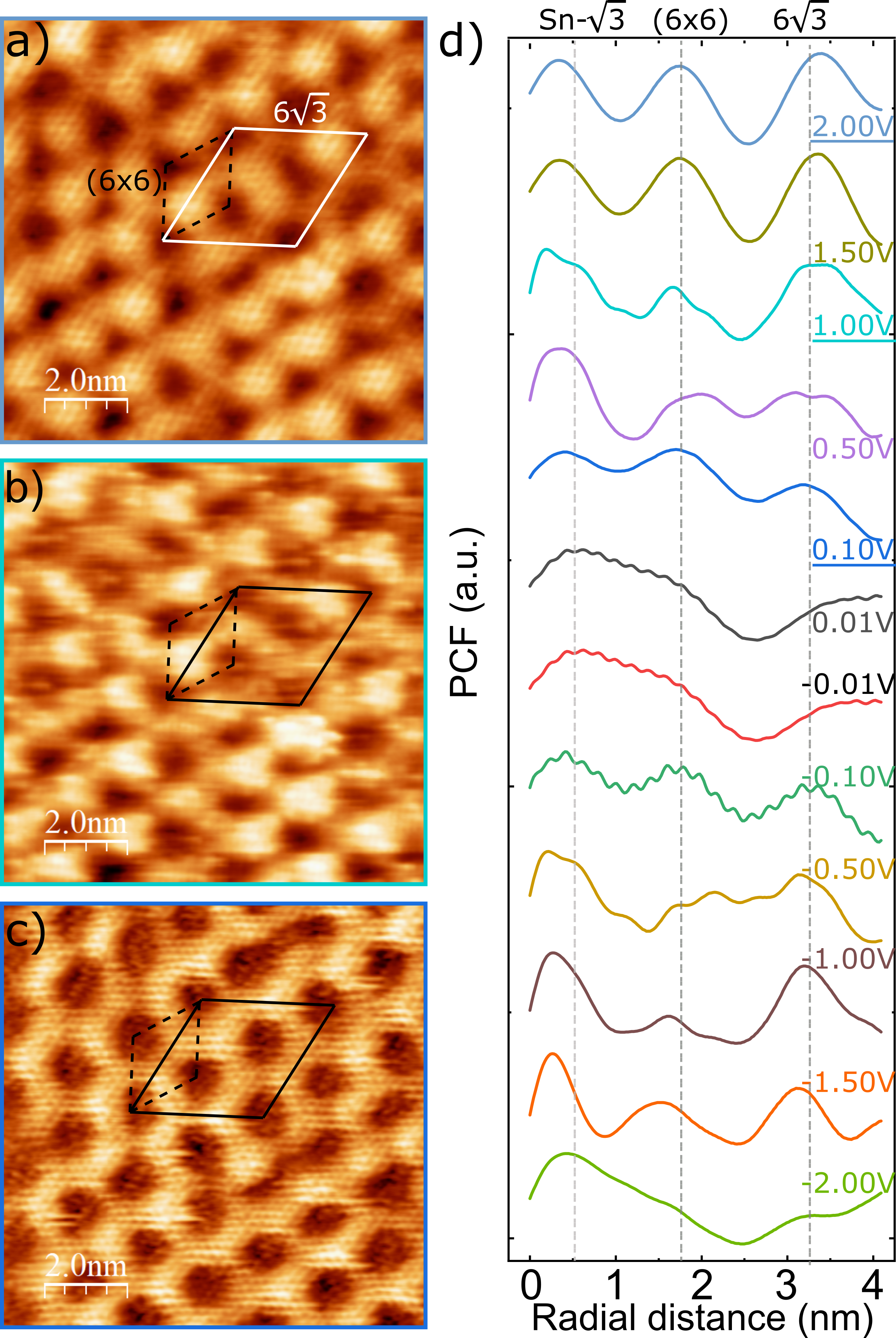}
	\caption{(a-c) The STM area phase of the green box in  Fig.~\ref{FIG2} at different bias conditions: (a) +2~V, (b) +1~V, (c) +0.1~V. (d) Pair correlation function (PCF) calculated from STM images taken at different bias voltages as indicated. The solid vertical lines show the distances of the main symmetries. At -0.5~V and +1~V bias voltages, distances belonging to $\sqrt3$a$_{SiC}$=5.3~\AA~ appear most clearly. The dashed and solid rhombuses denote the (6$\times$6) and  ($6\sqrt{3} \times 6\sqrt{3}$) unit cells, respectively.}\label{FIG3}
\end{figure}

In Fig.~\ref{FIG3} we show details of the phase which we associate with the Sn-($\sqrt{3} \times  \sqrt{3}$) phase. The images taken at different bias voltages clearly show different contrasts, i.e., various states at the interface are involved in the tunneling process.  While at +2~V mainly the SiC(1$\times$1) lattice with the (6$\times$6) symmetry is obvious, at small voltages close to the Fermi energy the graphene lattice is seen (Fig.~\ref{FIG3}c). At 1~V a high  contrast is achieved where the protrusion apparently follows the former bonds and reveals the (6$\times$6) superperiodicity.

In order to obtain a deeper insight into the underlying symmetries, we calculated the radial pair correlation functions from STM images taken at the same area but for different tunneling voltages ranging from +2~V to -2~V, shown in panel d).
As shown, various distances belonging to different symmetries of the interface are found. Besides the distances belonging to the (6 $\times$ 6) and ($6\sqrt{3} \times 6 \sqrt{3}$) symmetries, in particular for small bias voltages, the graphene lattice becomes visible (e.g. see the high frequency component at -0.1~V). Distances referring to the $\sqrt{3}$-symmetry show up only for two distinct voltages, namely -500~mV and +1000~mV. For these tunneling conditions also the second order of the $\sqrt{3}$ distance is visible. We will show in the following that these states are related to Sn, thus the Sn forms locally a $\sqrt{3} \times \sqrt{3}$ reconstruction, which was shown on SiC(0001) to host Mott states \cite{Glass2015}. 

Apparently, the adsorption sites of Sn atoms in the diluted phase are determined by the previous bonding sites of the buffer layer. While the saturation of these former BL bonds is sufficient to form freestanding graphene, the Sn coverage is insufficient to saturate all dangling bonds of the SiC surface. Consequently, this diluted phase inevitably should result in a n-type doped graphene layer, which seems to be a prerequisite to stabilize in turn  the Hubbard bands of the Sn structure.

\subsection{Electronic structure: EELS and STS}
\begin{figure}[t]
	\centering
	\includegraphics[width=.6\linewidth]{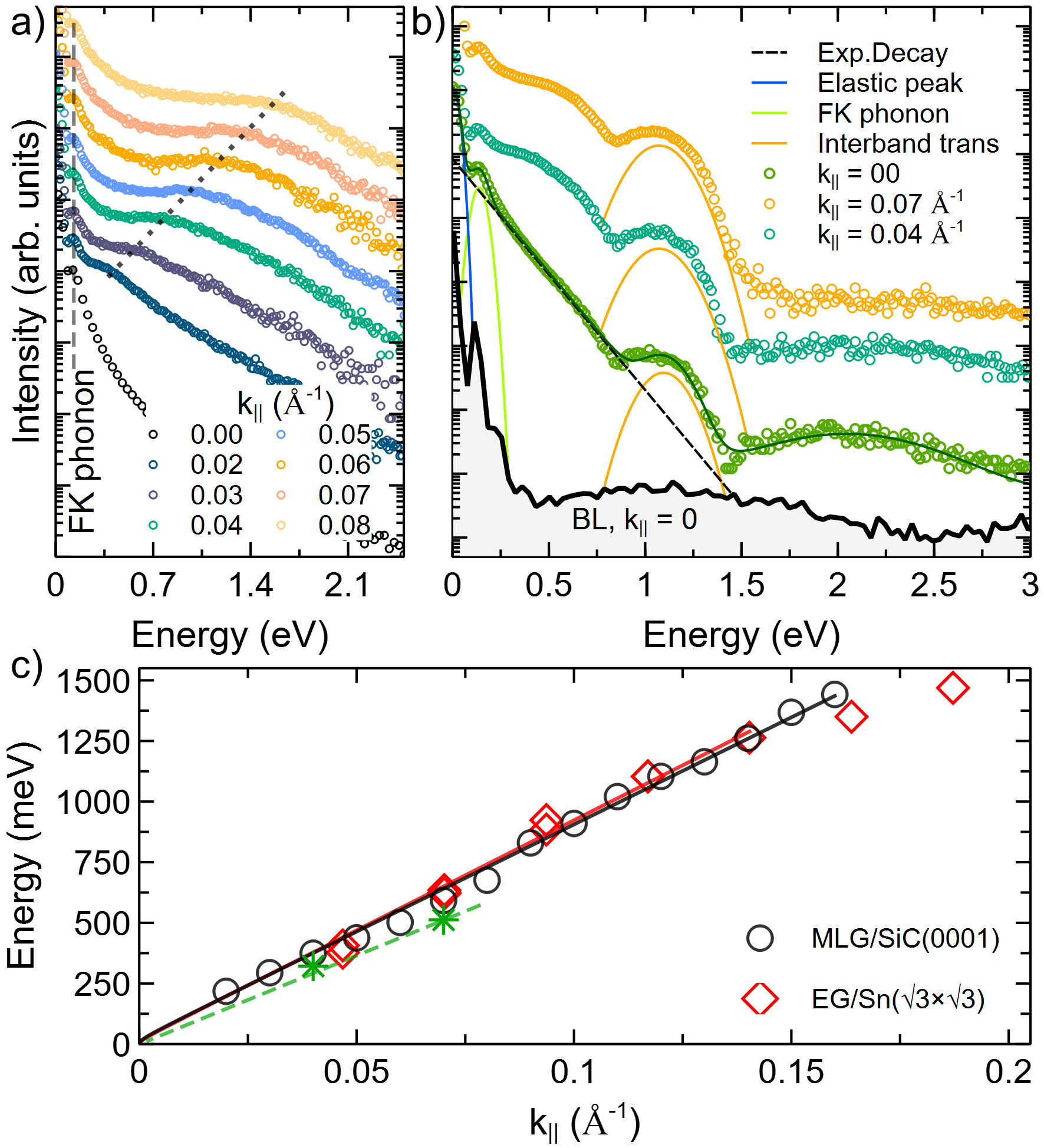}
	\caption{EEL spectra of the Sn-$\sqrt{3}$ phase after different steps of annealing. a) Loss spectra taken after the first appearance of $\sqrt{3}$-reflexes in LEED, showing the dispersing graphene sheet plasmon  loss \cite{Mamiyev2024}. b) Loss spectra after various heating cycles revealing besides the graphene plasmon an intense but non-dispersing loss-peak at around 1.2~eV. For reference, also the loss spectrum of the BL is shown. c) Sheet plasmon dispersions of the graphene plasmons deduced from  the spectra in a) and b). For reference, also the dispersion for the sheet plasmon in monolayer graphene (MLG) is shown.  The experiments were taken at 300~K with a primary electron energy of 20~eV. }\label{FIG4}
\end{figure}

The aforementioned phases were further analyzed by means of electron energy loss spectroscopy. Figure \ref{FIG4} displays EEL spectra of nominally Sn-$\sqrt{3}$ phases. Panel a) reveals a sheet plasmon loss peak that  disperses with increasing the momentum transfer. The analysis of this collective mode indicates that this plasmon feature corresponds to n-type doped graphene, with a doping level of approximately n=7.8 $\times$ 10$^{12}$ cm$^{-2}$ carriers, shifting the Fermi level around $E_{\mathrm{F}}=330$~meV above the Dirac point \cite{Mamiyev2024}. The dispersion is plotted in Fig.~\ref{FIG4}c) 

Upon further heating cycles of the Sn-$\sqrt{3}$ phase  the LEED intensity of the $\sqrt{3}$-reflexes become more intense (cf. Fig.~\ref{FIG1}e) and  the loss spectrum underwent significant changes, as shown in Fig.~\ref{FIG4}b). Besides the sheet plasmon, visible at lower loss energies, the spectrum shows a distinct non-dispersing loss peak at 1.2 eV. The high loss intensity of this peak suggests also a high combined density of states, e.g., as for Mott-Hubbard bands.  The loss energy is in reasonable agreement with the extrapolation from the position of the LHB seen in photoemission for the triangular Sn lattice on SiC(0001) \cite{Glass2015}. The fitting of the spectra reveals that the loss peak is astonishingly broad (FWHM=0.4~eV). For reference, we also present the loss spectrum of a buffer layer (BL), demonstrating that while the Sn-$\sqrt{3}$ phase shares the same symmetry as an ordinary BL, it exhibits distinct electronic properties.
As mentioned, there is also  the graphene plasmon at lower loss energies visible. The remaining dispersing nature of this state is highlighted through spectral fitting and its  dispersion is plotted also in panel c). From the similarity of both we conclude that freestanding graphene is indeed present and that also the doping has not severely changed.  Therefore, in agreement  with  the evolution of the intensity of the $\sqrt{3}$ intensity shown in Fig.~\ref{FIG1}e), also EELS shows that the  Sn-$\sqrt{3}$ domains become gradually larger by subsequent annealing.

\begin{figure}[t]
	\begin{center}
	 \includegraphics[width= .5\linewidth]{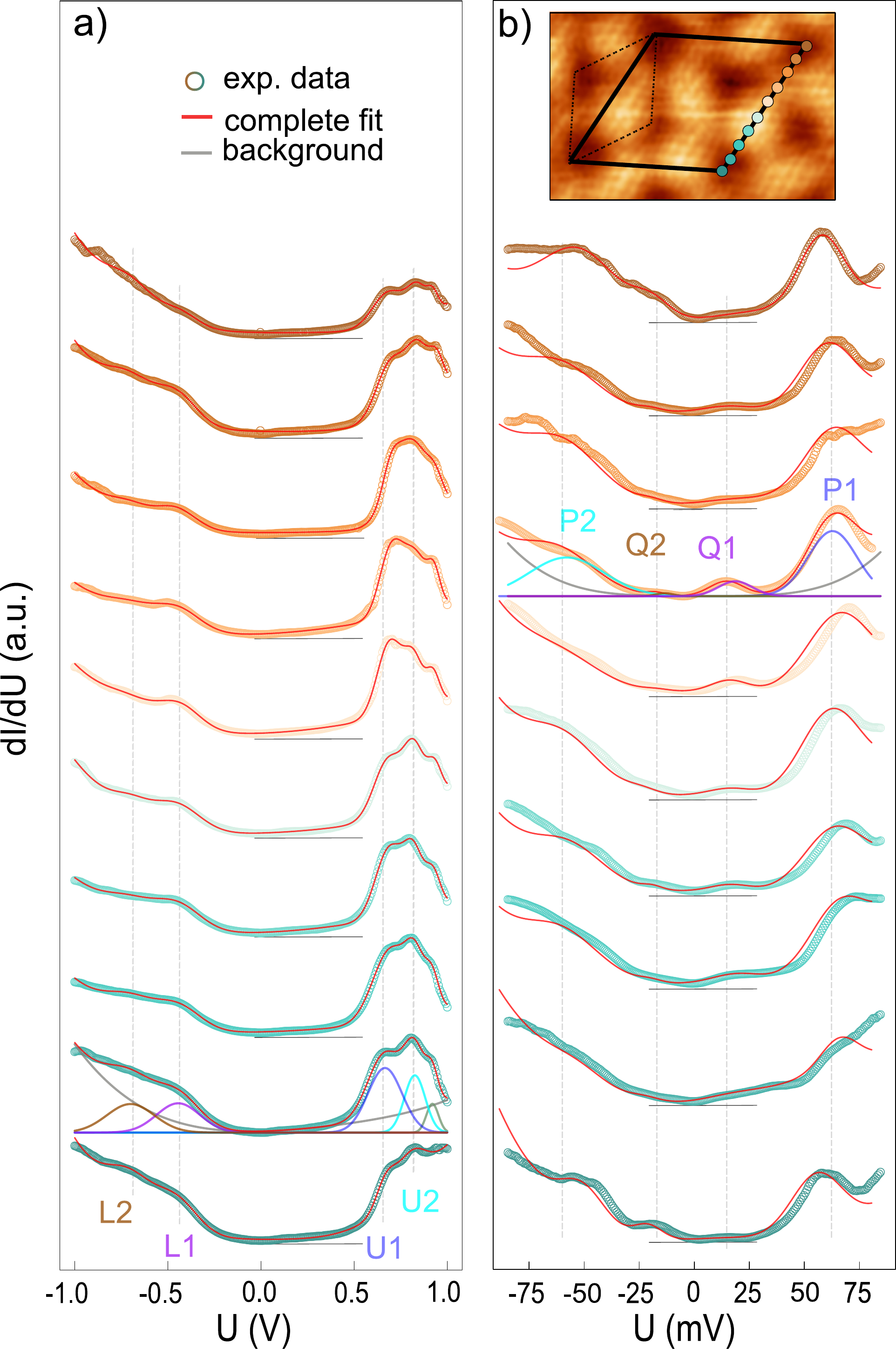}
		\caption{High resolution STS spectra taken along the unit cell of the $6\sqrt{3}$ unit cell shown in the inset. (a) Voltage range from -1 to+1~V (set point +2~V, 500~pA). The color code refers to the positions shown in the inset.  (b) Same sequence as in  (a), but with a voltage range from -0.1 to +0.1~V (set point  +0.1~V, 500pA). The detailed fit model for the two regimes are shown exemplarily.}\label{FIG5}
	\end{center}
\end{figure}

\begin{figure}[t]
	\begin{center}
		\includegraphics[width= .45\linewidth]{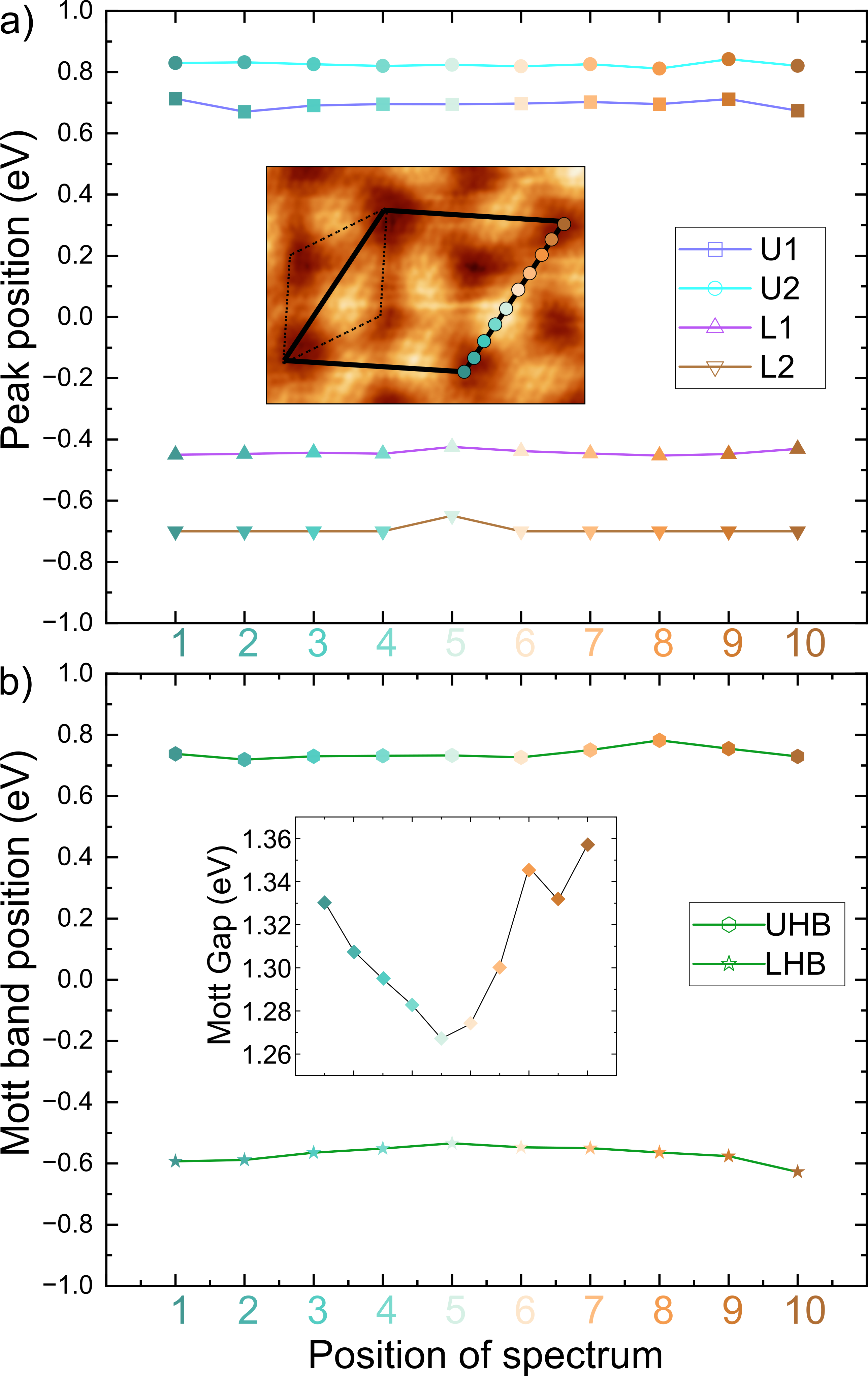}
		\caption{Evaluation of the STS spectra shown in Fig.~\ref{FIG5}. (a) Peak positions of the lower and upper Hubbard band peaks, U1, U2, L1 and L2, across the unit cell as marked in the inset for the peaks discussed in context of Fig.~\ref{FIG5}a). (b) Averaged peak positions of the U1, U2, L1 and L2 states. The inset shows the difference, i.e., the variation of the Mott gap across the unit cell.}\label{FIG6}
	\end{center}
\end{figure}

High resolution dI/dV spectra  taken along the $6\sqrt{3}$ unit cell in the range from -1~V to +1~V are shown in Fig.~\ref{FIG5}a). Most obvious are the peaks at -0.5~V and around +0.7~V. The size of this gap fits nicely to the energy loss of 1.2~eV found in EELS (cf. Fig.~\ref{FIG4}b). Each of the peaks consists of  sub-peaks, respectively, as deduced from the fitting of the data. The sub-peak structure is for example shown for one spectrum. Moreover, the substructure of the initial and final states seems to be responsible for the large peak width of around 0.3~eV.  
The detailed analysis of the peaks by our fitting routine revealed that the peak at around -0.5~eV is split into two peaks located at -430~meV (L1) and -680~meV (L2). Also the unoccupied states  at +0.7~eV reveals a substructure U1, U2.  In view of the DMFT calculations presented hereafter, these states seem to resemble the lower and upper Hubbard bands (LHB, UHB). As obvious, the energetic positions of the peaks and their intensity show some slight variation along the axis of the (6$\sqrt{3} \times $6$\sqrt{3}$) unit cell. Details will be discussed in context of  Fig.~\ref{FIG6}a). The STS measurements revealed splitting within the Hubbard bands, which is consistent with the DMFT calculations (see below).

In Fig.~\ref{FIG5}b) high resolution STS spectra around the Fermi level are shown. Details were again deduced from fitting the data, as exemplarily for one spectrum. Within this energy window mainly four peaks are seen. The outer ones (P1,P2)  at around $\pm60$~meV define a gap-like feature around the Fermi level arising from a suppression of electronic tunneling to graphene states near E$_F$.  At around 60~meV, an enhancement of electronic tunneling sets in  due to a phonon-mediated inelastic channel \cite{Zhang2008}. The presence of this gap is a hallmark for freestanding and defect-free graphene. Moreover, unlike the Sn(1$\times $1)  patches, we found also no  Kekul\'{e}-distortion in the Sn-$\sqrt{3}$ phases, thus we can assume here a comparably small hybridization between the graphene and Sn interface layer.

Moreover, the sequence of spectra show peaks close to E$_F$ at Q2=-18~meV and Q1=14~meV. Most likely,  these spectroscopic features hint towards signatures of a correlated metal due to the interplay between hybridization between the Sn-layer and graphene, which can be partly reproduced by DMFT, however, by assuming a stronger hybridization or charge transfer.   

We have conducted a detailed analysis of the Sn-states. Figure \ref{FIG6}a) shows peak positions of the  L1, L2, and U1, U2 states along the unit cell, as indicated in the inset. Including also the different intensities for each at the peaks and each of the position, we determined the average position of the lower and upper Hubbard band, shown in panel b). The bands and correspondingly the Mott gap, shown as inset, vary across the unit cell. In context of the theoretical results, presented hereafter, T4- and H-positions of the Sn atom with respect to the graphene lattice (cf. Fig.~\ref{FIG0}) gives rise to different hybridization strength, thus to a variation of the Hubbard bands.  The overall variation is in the order of 0.1~eV explains the  comparably large peak width seen in EELS (cf. Fig.~\ref{FIG4})  and agrees well our theoretical findings. and

\begin{figure*} [tb]	 
	\includegraphics[width=.8\textwidth]{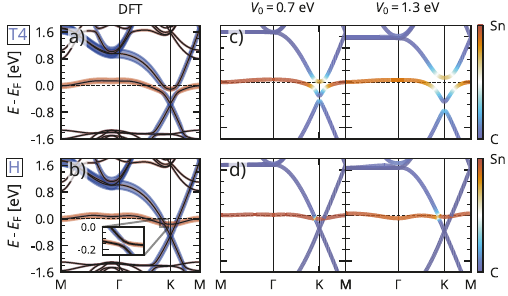}
	\caption{(a,b): Electronic structure from DFT with flatband highlighting of the orbital weight by Sn (orange) and graphene (blue) $p_z$ orbitals for the T4 (a) and H (b) cell (cf.~Fig.~\ref{FIG0}. The inset in b) shows a closeup of the hybridization gap along the $\Gamma$K direction. (c,d): Model band structure at $V_0=0.7$~eV and 1.3~eV without local Coulomb interaction ($U=0$). The color bar indicates the spectral weight of orbital $m\in\lbrace p_z^{\mathrm{Sn}},p_z^{\mathrm{C}}\rbrace$ in the lowest-lying non-interacting band.}
	\label{fig_t1}
\end{figure*}

\subsection{Interplay of hybridization, electronic correlation and doping: DFT and DMFT}

To understand qualitatively how the low-energy electronic structure of graphene is affected by the hybridization with Sn, we used a simplified $\sqrt{3}\times\sqrt{3}$ supercell with different graphene alignments (T4 and H positions) shown in Fig.~\ref{FIG0} accounting for the experimental findings discussed in the context of Figs.~\ref{FIG1}--\ref{FIG3}. 

The corresponding band structures from DFT are shown in Figure~\ref{fig_t1}a) and b) with fatbands of Sn and graphene $p_z$ orbitals. Both stacking configurations show a flat band at the Fermi level made up by Sn-$p_z$ orbitals coexisting with electron doped Dirac bands of graphene. Such a situation is reminiscent of the typical band structures of heavy fermion materials including $f$-electron compounds~\cite{Stewart1984,Checkelsky2024}, magic-angle twisted graphene multilayers~\cite{Bernevig2024,Song2022,Rai2024}, and also transition metal dichalcogenide multilayer structures~\cite{crippa2024}. Close to the crossing of the two kinds of bands around the K-point, we find hybridization induced avoided crossings in the T4 structure, which are largely suppressed in the H structure. This stacking induced modulation of the hybridization arises from the following selection rule: the graphene states at $K$ transform according to the $E_1$ and $E_2$ representations under $C_{6v}$ symmetry with respect to the hexagon centers, whereas Sn $p_z$ orbitals belong to the $A_1$ representation~\cite{wehling_orbitally_2010}. This kind of selection rule does not apply in the T4 position.

Figures~\ref{fig_t1}c) and d) present the band structures obtained from solving the tight-binding Hamiltonian $H_0$ (Eq.~(\ref{eq:model})) for $\Delta=0.5$~eV and two representative hybridization strengths of $V_0=0.7$~eV and $V_0=1.3$~eV neglecting correlation effects.
The aforementioned selection rules manifests also here: when comparing the band structures with $V_0=0.7$ and $V_0=1.3$~eV in the T4 structure, we see an enhancement of the corresponding hybridization gaps, while band structures for the H structure are almost insensitive to this change in $V_0$. In comparison, the tight-binding bands shown in Figs.~\ref{fig_t1}c) and d) closely replicate the corresponding DFT bands around the Fermi level for each structure presented in Figs.~\ref{fig_t1}a) and b). Thus, $V_0=0.7$ and $1.3$~eV should fall into a reasonable range of $V_0$.

\begin{figure*} [t]	 
	\includegraphics[width=0.9\textwidth]{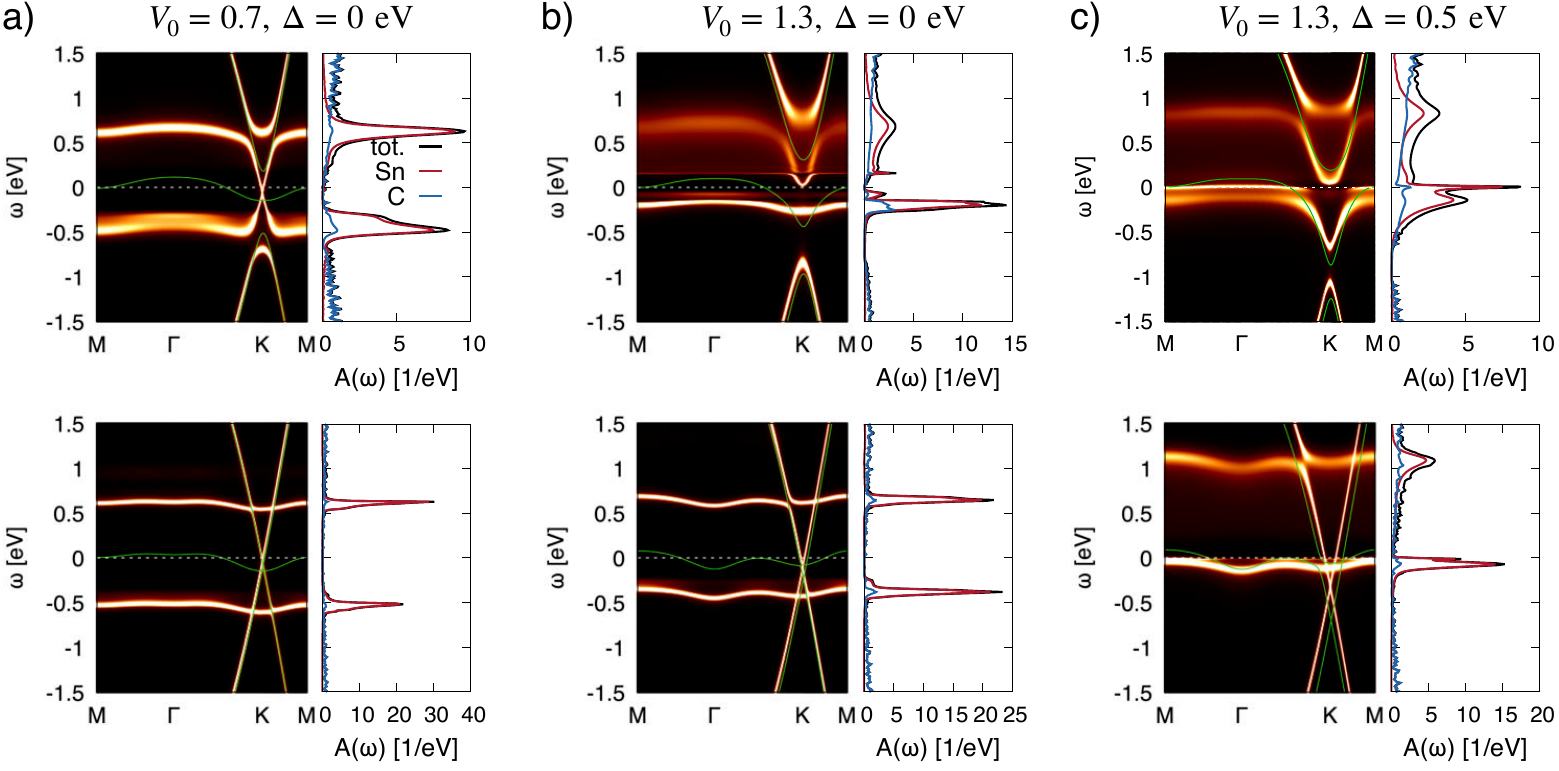}
	\caption{(a--c) The momentum-dependent spectral function $A(\mathbf{k},\omega)$ (color coded) and the orbital-resolved momentum-integrated spectral function $A(\omega)$ obtained from DMFT for $U=1.2$~eV. Top panels: Sn in the T4 structure. Bottom panels: Sn in the H structure. The chemical potential is at $\omega=0$. The corresponding non-interacting bands (green solid lines) are overlaid. }
	\label{fig_t2}
\end{figure*}

\textit{DMFT results:}
We now investigate how the Hubbard interaction modifies the electronic excitation spectra. Figures~\ref{fig_t2}(a--c) show the momentum-dependent spectral function $A(\mathbf{k},\omega)$ and the momentum-integrated spectral function $A(\omega)$, which can be measured by photoemission or scanning tunneling spectroscopy, respectively. 

The most notable and robust feature is the emergence of the upper Hubbard band (UHB) and the lower Hubbard band (LHB) from the flat Sn-$p_z$ derived band for both T4 and H structures. This shows that regardless of the model details and correspondingly the local configuration of graphene, the Sn $p_z$ orbitals are in a regime of strong Mott-Hubbard type correlations.

While strong correlations in the Sn $p_z$ states are generic here, the stacking configurations (namely, T4 vs.~H), the hybridization strength $V_0$, and the energy offset $\Delta$ can strongly affect the nature of electronic states in the Sn-graphene hybrid system. Due to suppressed hybridization near the Dirac point, the Hubbard bands are very sharp and closely resemble an atomic limit for the H structures at $\Delta=0$ [lower panels in Fig.~\ref{fig_t2}(a,b)]. In the T4 states, hybridization broadens the Hubbard bands [upper panels in Fig.~\ref{fig_t2}(a,b)].

Interestingly, the spectra shown in Fig.~\ref{fig_t2}(a,b) for weak to intermediate hybridization naturally explain several of our experimental findings. The ``Mott gaps'' (the energy level difference between UHB and LHB) on the order of $1$~eV fit to the loss energy in EELS [cf. Fig.~\ref{FIG4}b)] and match the gap between the L1/L2 and U1/U2 states in STS [cf. Fig.~\ref{FIG5}(a)], resembling the LHB and UHB, respectively. As already shown, the loss peak is comparatively broad. In addition, the STS measurements reveal a splitting of the Hubbard bands, which is according to the calculated spectra shown in Fig.~\ref{fig_t2}. This can be understood as a hybridization effect --- particularly in the T4 configurations [cf. upper panels in Fig.~\ref{fig_t2}a) and b)]. The splitting of Hubbard-type bands through hybridization effects is quite common in correlated multiband systems as exemplified by the Zhang-Rice singlet from cuprate high-$T_\mathrm{c}$ superconductors~\cite{Zhang_Rice_PhysRevB88,weber_2010,Werner_PhysRevB15}.

Potential energy offsets directly affect the position of the Hubbard bands, leading from an (almost) particle-hole symmetric situation at $\Delta\to 0$ [Fig.~\ref{fig_t2}(a,b)] towards a ``mixed valence'' case where the lower Hubbard band is pinned to the Fermi level [Fig.~\ref{fig_t2}c)]. Comparison of calculated spectra to our STS experiments suggests $|\Delta|\ll 0.5$~eV.

One of the appealing aspects of the system at hand is that graphene offers the opportunity of two-side functionalization and correspondingly doping of the system~\cite{schumacher_backside_2013}. We thus study the case featuring stronger charge transfer closer [$\Delta=0.5$~eV, Fig.~\ref{fig_t2}(c)]:
From the Hubbard band pinned to the Fermi level, quasiparticle peaks emerge both in the T4 and in the H configuration.
Yet, the nature of the emergent state upon charge transfer is very different in each case. In the H structure, the Sn $p_z$ merely shift but hybridization remains suppressed as it must be due to symmetry reasons. Thus, we have essentially a doped Mott insulator coexisting with a charge reservoir (the Dirac electrons of graphene). This situation bears a close resemblance to 1T/1H Ta-dichalcogenide bilayers \cite{crippa2024} in which charge transfer from uncorrelated electrons (1H layer) to a Mott insulator (1T layer) gives rise to a correlated metal as observed in experiments.

On the other hand, in the T4 case the doped Mott state of the Sn $p_z$ band strongly hybridizes with the graphene Dirac electrons as indicated in the upper panel in Fig.~\ref{fig_t2}c). This case is thus closer to systems like magic-angle twisted bilayer graphene featuring similar hybridization effects \cite{Cao2018,Cao_2018_2,Song2022,Rai2024}.

\section{Summary and Conclusions}

The Sn-intercalated graphene system resembles a novel platform of proximitized Dirac and correlated electrons. In the $\sqrt{3}$-patches, Sn $p_z$ electrons exhibit robust correlation effects manifesting as characteristic Hubbard bands in STS and loss peaks in EELS. The system reveals spatially modulated hybridization between the Dirac and correlated electrons. Theoretical modeling and experiment show excellent agreement regarding the spectral properties and reveal symmetry related selection rules as origin of the modulated hybridization.

The correlated electron platform introduced here offers special tunability: Our analysis reveals that besides the hybridization and Coulomb interaction also the charge transfer plays an important role for the electronic state that emerges in these artificial correlated systems. While the former are defined by the specifics of the material, the charge transfer can be controlled in these epitaxial systems quite easily, e.g., by  molecular doping with F4-TCNQ. Further experiments are underway to investigate systematically the interplay of doped Mott states in direct spatial proximity to more hybridized heavy fermions.

\vspace{2em}
\textbf{Acknowledgement}
We gratefully acknowledge financial support from the DFG through FOR5242 (TE386/22-1 and WE5342/7-1). The DMFT calculations were done on the supercomputer Lise at NHR@ZIB as part of the NHR infrastructure under the project hhp00056.

\appendix


\end{document}